\documentclass[preprint,superscriptaddress,showpacs,amsmath,amssymb,aps,pra]{revtex4-2}
\usepackage{pifont}
\usepackage{times}
\usepackage{amssymb}
\usepackage{graphicx}
\usepackage{dcolumn}
\usepackage{dsfont}
\usepackage{bm}
\usepackage{subfigure}
\usepackage[ruled]{algorithm2e}
\usepackage{hyperref}
\usepackage{appendix}
\usepackage{lipsum}

\newcommand{\T}{\mbox{$\mathrm{tr}$}}

\newcommand{\ket}[1]{\ensuremath{| #1 \rangle}}

\hypersetup{colorlinks=true, citecolor=blue, urlcolor=blue, linkcolor=blue}

\newtheorem{theorem}{Theorem}

\newtheorem{lemma}{Lemma}

\input amssym.def

\begin{document}
\title{Tetrahedron genuine entanglement measure of four-qubit systems}
\author{Meng-Li Guo}
\email{guoml@cnu.edu.cn}
\affiliation{School of Mathematical Sciences, Capital Normal University, Beijing 100048, China}
\author{Zhi-Xiang Jin}
\affiliation{School of Computer Science and Technology, Dongguan University of Technology, Dongguan, 523808, China}
\author{Bo Li}
\affiliation{School of Computer and Computing Science, Hangzhou City University, Hangzhou 310015, China}
\author{Shao-Ming Fei}
\email{feishm@cnu.edu.cn}
\affiliation{School of Mathematical Sciences, Capital Normal University, Beijing 100048, China}
	
\begin{abstract}
Quantifying genuine entanglement is a key task in quantum information theory. We study the quantification of genuine multipartite entanglement for four-qubit systems.
Based on the concurrence of nine different classes of four-qubit states, with each class being closed under stochastic local operation and classical communication, we construct a concurrence tetrahedron. Proper genuine four-qubit entanglement measure is presented by using the volume of the concurrence tetrahedron. For non genuine entangled pure states, the four-qubit entanglement measure classifies the bi-separable entanglement. We show that the concurrence tetrahedron based measure of genuine four-qubit entanglement is not equivalent to the genuine four-partite entanglement concurrence. We illustrate the advantages of the concurrence tetrahedron by detailed examples.
\end{abstract}
	
\maketitle
	
	
\section{Introduction}
Quantum entanglement is an essential feature of quantum mechanics that distinguishes the quantum from the classical world \cite{op1,op2,op3,op4,op5,op6}.
It has potential applications in quantum information processing such as quantum cryptography \cite{op7}, quantum dense coding \cite{op8},  quantum secret sharing  \cite{op9}, quantum teleportation \cite{op10} and measurement-based quantum computing \cite{op11}.
To use the entanglement as a resource is not only a matter of how to detect it, but also how to quantify it. The entanglement measures for bipartite systems have been well studied \cite{op12,op13,op14}, such as the concurrence \cite{op13,op16,Con}, entanglement of formation \cite{op18,op19} and negativity \cite{op20,Neg}. For a bipartite pure state
$|\psi\rangle_{AB}$ in a finite-dimensional Hilbert space $ \mathcal{H}_A\otimes \mathcal{H}_B=\mathbb{C}^{d_1}\otimes\mathbb{C}^{d_2}$ the concurrence is defined as \cite{Con} $C( |\psi \rangle_{AB})=\sqrt{2[1-Tr(\rho^{2}_A)]}$, where $\rho_A$ is the reduced density matrix by tracing over the subsystem $B$, $\rho_A=Tr_B(|\psi\rangle_{AB}\langle\psi|)$.
The concurrence is extended to mixed states $\rho=\sum\limits_{i} p_i | \psi_i \rangle \langle \psi_i |$, $0 \leq p_i \leq 1$, $\sum\limits_{i} p_i =1$, by the convex roof extension
$C(\rho_{AB})=\min\limits_{\left\lbrace {p_i, | \psi_i \rangle}\right\rbrace}  \sum\limits_{i} p_i C(| \psi_i \rangle )$, where the minimum is taken over all possible pure state decompositions of $\rho_{AB}$.
	
For entanglement in multipartite systems, although the experimental observation has been successfully implemented \cite{op22,op23,op24,Ma}, the quantification of multipartite entanglement is still far from being satisfied. In \cite{Ma} Ma et al. introduced the genuine multipartite entanglement measure.
A measure of genuine multipartite entanglement (GME) $E(\rho)$ of a state $\rho$ should at least satisfy: (a) the measure must be zero for all product and biseparable states; (b) the measure must be positive for all non-biseparable states (e.g., the Greenberg-Horn-Selinger (GHZ) state and W state for the three-qubit case). In addition, besides conditions (a) and (b), a well defined GME measure also needs to satisfy the following condition: (c) the measure should be nonincreasing under local operations and classical communications.
Recently, Xie et al. \cite{xie} proposed a new triangle measure specifically for tripartite systems, which has a simple form and an elegant geometric interpretation, but does not satisfy the GME requirement (c). In \cite{JZX} the authors presented an improved GME measure that satisfies all the requirements. Further related researches have also been presented very recently \cite{Gyu,Dai,Li,Pul}.
	
In this paper, we analyze in detail the multipartite entanglement measures for four-qubit systems and show how the concurrence tetrahedron constitute a well defined GME measure. The rest of this article is organized as follows. In Sec. \uppercase\expandafter{\romannumeral2}, we analyze and calculate the concurrences of all nine-type four-qubit states. It is found that except for the product states and biseparable states, the non-biseparable states form a concurrence tetrahedron, which shows that the concurrence tetrahedron is a new bona fide entanglement measure. In addition, we also compare the concurrence tetrahedron with genuinely multipartite concurrence (GMC). It is found that GMC and the concurrence tetrahedron are not equivalent. For four-qubit systems the concurrence tetrahedron has more advantages than the generalized method of moments. Finally we discuss and summarize the results in Sec.\uppercase\expandafter{\romannumeral3}.
	
\section{Tetrahedral entanglement measurement of four-qubit states}
Two quantum states that can be converted to each other with limited success probability under local operation and classical communication (LOCC) or stochastic local quantum operations assisted by classical communication (SLOCC) belong to the same entanglement class \cite{Sci,Sci5,Sci6}. For three-qubit systems, there are two types of entangled states with completely different properties: the GHZ states and W states. Westerlet et al. show that there are nine different types of entanglements for four-qubit systems \cite{nine} under SLOCC operations.
For an arbitrary $N$-qubit pure state $|\Psi\rangle$, denote $E_{j}=E_{j|1...k\neq j...N}$ the entanglement under the bipartition of the $j$-th qubit and the rest ones. It is shown that \cite{Qian,LMX}, $E_{j} \leq  \sum_{k\neq j}E_{k}$, which is valid for a generic entanglement measure $E$ such as the von Neumann entropy \cite{Von}, concurrence \cite{Con} and negativity \cite{Neg}.

\begin{figure}
	\scalebox{1.2}{\includegraphics[width=10cm]{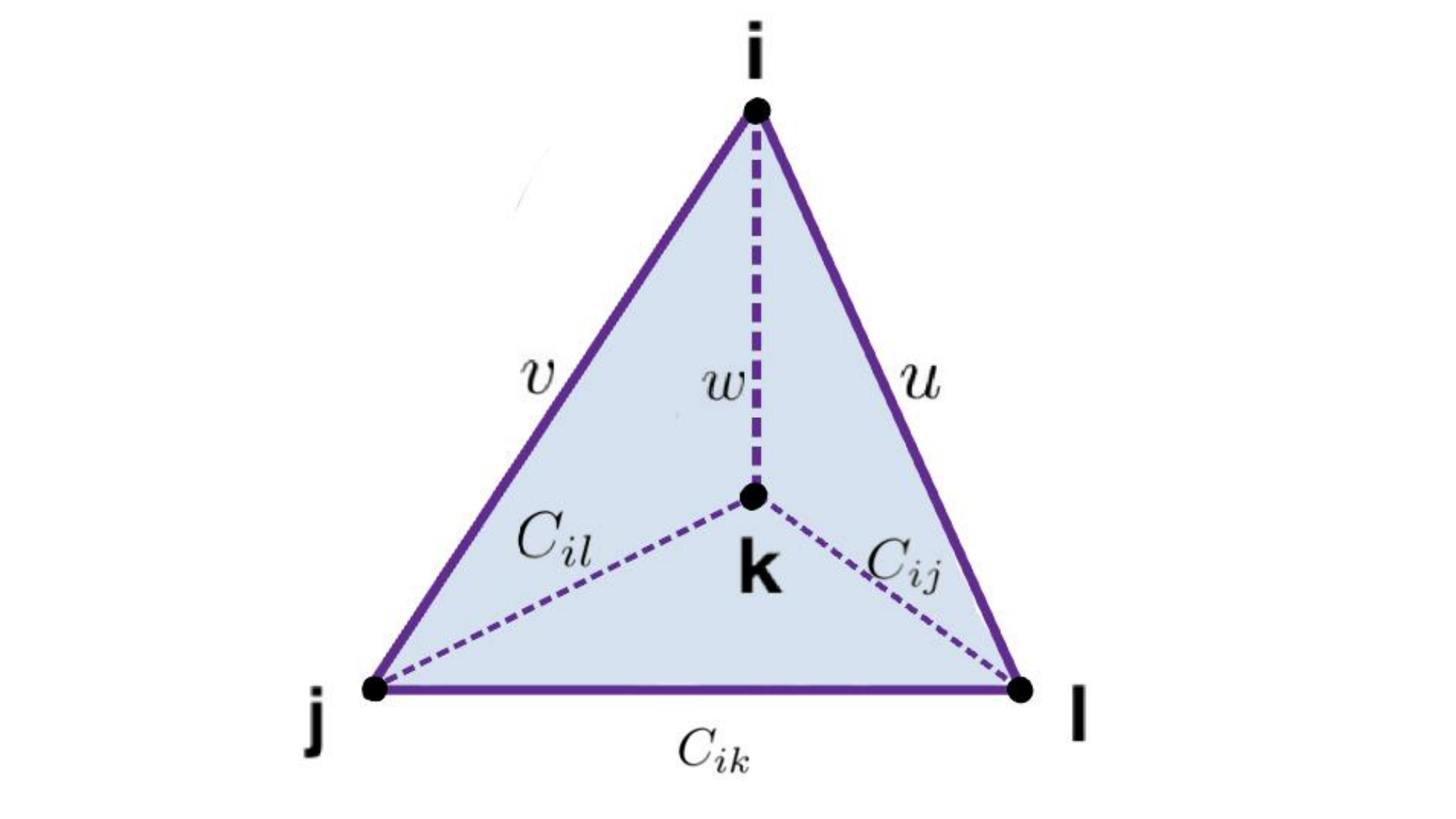}}
	\caption{The concurrence tetrahedron of a four-qubit system. $u$, $v$ and $w$ are the smallest three of $C_{i}$, $C_j$, $C_k$ and $C_l$, which are the edges of the tetrahedron. $C_{ij}, C_{ik}$ and $C_{il}$ are the sides of the base triangle.}
	\label{Fig1}
\end{figure}

Consider the concurrence of general four-qubit pure states $|\psi_{ABCD} \rangle$. Denote $C_{i}=C_{i|jkl}$ and $C_{ij}=C_{ij|kl}$ the concurrences under bipartitions $i|jkl$ and $ij|kl$, respectively, where $i\neq j\neq k\neq l\in\{1,2,3,4\}$. We have
\begin{equation}\label{inq1}
C_{i|jkl}\leq C_{j|kli}+C_{k|ijl}+C_{l|ijk}
\end{equation}
and
\begin{equation}\label{inq2}
C_{ij|kl}\leq C_{ik|lj}+C_{il|jk},
\end{equation}
where the equality $C_{ij|kl}= C_{ik|lj}+C_{il|jk}$ holds when the state is biseparable under the partition \cite{Gyu,Qian}. To have an explicit geometric picture, consider the value of each concurrences $C_{i}$ and $C_{ij}$ as the length of lines.
For non-biseparable states, according to inequality (\ref{inq2}), $C_{ij}$, $C_{ik}$ and $C_{il}$ form a triangle. Moreover, let $u,v,w$ be the smallest three of $\{C_{i}, C_j, C_k, C_l\}$ with $0\leq u\leq v \leq w$. Then $u,v,w$ and $C_{ij}$, $C_{ik}$, $C_{il}$ form a tetrahedron, as shown in Fig. \ref{Fig1}. We call it {\it concurrence tetrahedron}.

We first give a lemma that will be used in deriving our main results.
	
\begin{lemma}\label{lemma1}
Let $\triangle ABC$ be a triangle with three vertices $A$, $B$ and $C$. Denote $3R$ the sum of the distances from the center $O$ of the circumscribed circle of the triangle to the three vertices. Let $D$ is a point which is not coplanar with the triangle $\triangle ABC$, and $H$ is the sum of the distances from $D$ to the three vertices of $\triangle ABC$. If $H-3R>0$, then $A$, $B$, $C$ and $D$ can form a tetrahedron.
\end{lemma}
	
{\sf Proof}
Since the sum of the distances from the center of the circumscribed circle of the triangle to the three vertices of the triangle is the longest, if $H(|\psi_{ABCD} \rangle) \leq 3R$, then $D$ is inside the triangle $\triangle ABC$, and the four points $A,B,C,D$ cannot form a tetrahedron. Particularly, when $H(|\psi_{ABCD} \rangle)=3R$, $D$ is the center of the circumscribed circle of the triangle. Therefore, $H(|\psi_{ABCD} \rangle)>3R$ implies that $D$ is outside the $\triangle BCD$, and the $A,B,C,D$ can form a tetrahedron. $\Box$
	
\begin{figure}
\includegraphics[width=15.3cm]{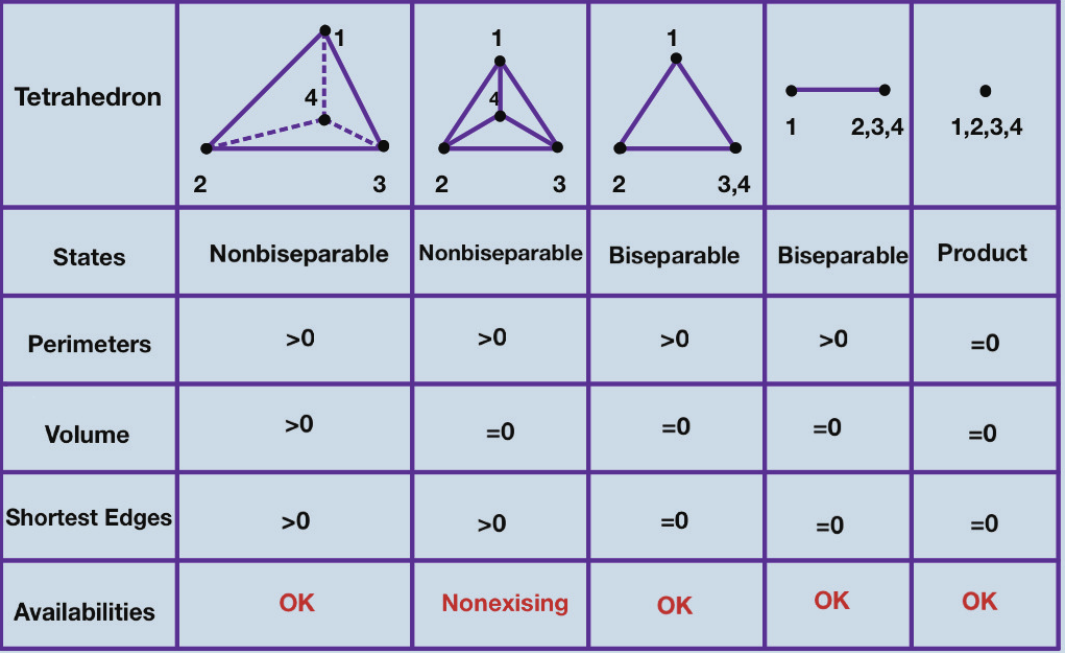}
\caption{Four-qubit entanglement and concurrence tetrahedron: ``Perimeter" refers to the sum of all the edges of a tetrahedron. ``Shortest side" refers to the length of the shortest edge of a tetrahedron. ``Non-existence" means that there is no four-point coplanar case, namely, there is no situation such that the tetrahedron degenerates into a planar quadrilateral, which is from the analysis of the nine kinds of states in the proof of Theorem 1.}
\label{Fig3}
\end{figure}

The volume of the concurrence tetrahedron is another interesting quantity. For both product and biseparate states, the lengths of the tetrahedron is zero, i.e., the volume is zero and thus the condition (a) of GME is satisfied. In addition, the volume of the concurrence tetrahedron also satisfies the condition (b).
In Fig. \ref{Fig3} we list in table the relations between the entanglement of four-qubit states and the corresponding concurrence tetrahedron.
	
\begin{theorem}\label{Thm1}
For any four-qubit pure state $|\psi_{ABCD} \rangle$, the volume of the concurrence tetrahedron defines a bona fide GME measure,
\begin{equation*}
\mathcal{V}_{1234}(|\psi_{ABCD} \rangle)= \frac{1}{12}\sqrt{4 u^2 v^2 w^2-u^2 D^2- v^2 E^2-w^2 F^2+DEF},
\end{equation*}
where $D=v^2+w^2-C_{ij}^2$, $E=u^2+w^2-C_{ik}^2$, $F=u^2+v^2-C_{il}^2$, $u,v,w$ and $C_{ij}$, $C_{ik}$, $C_{il}$ are the edges of concurrence tetrahedron.
\end{theorem}
	
{\sf Proof}
We first prove that $|\psi_{ABCD}\rangle$ is GME iff $\mathcal{V}_{1234}(|\psi_{ABCD} \rangle)>0$. On one hand, if $\mathcal{V}_{1234}(|\psi_{ABCD} \rangle)>0$, then each edge of the concurrence tetrahedron, $u$, $v$, $w$, $C_{ij}$, $C_{ik}$ and $C_{il}$ are all greater than zero. Hence $|\psi_{ABCD} \rangle$ is a genuine multipartite entangled state. On the other hand, if $\mathcal{V}_{1234}(|\psi_{ABCD} \rangle)=0$, at least one edge of the concurrence tetrahedron is zero, namely, either $\psi_{ABCD}=\rho_{A}\otimes \rho_{BCD}$ or $\psi_{ABCD}=\rho_{B}\otimes \rho_{ACD}$ or $\psi_{ABCD}=\rho_{C}\otimes \rho_{ABD}$ or $\psi_{ABCD}=\rho_{D}\otimes \rho_{ABC}$. Therefore, $\psi_{ABCD}$ is not a genuine multipartite entangled state.
	
We next prove that $\mathcal{V}_{1234}(|\psi_{ABCD} \rangle)$ cannot increase under LOCC, i.e., $\mathcal{V}_{1234}(|\psi_{ABCD} \rangle)\geq \mathcal{V}_{1234}(\Lambda(|\psi_{ABCD} \rangle))$ for any LOCC map $\Lambda$. As concurrence $C$ does not increase under LOCCs, the edges of the tetrahedron do not increase under LOCCs. Hence, we only need to prove that the volume $\mathcal{V}_{1234}$ is an increasing function of its edges $u$, $v$, $w$, $C_{ij}$, $C_{ik}$ and $C_{il}$.
	
In the concurrence tetrahedron as shown in Fig. \ref{Fig1}, with respect to Lemma \ref{lemma1} we set $H(|\psi \rangle)=u+v+w$ and denote $R$ the radius of the circumscribed circle of the triangle on the bottom, $G(|\psi \rangle)=H(|\psi \rangle)-3R$, where $R=\frac{C_{ij} C_{ik} C_{il}}{4 \sqrt{p (p-C_{ij}) (p- C_{ik}) (p-C_{il})}}$ with $p=\frac{1}{2} (C_{ij}+C_{ik}+C_{il})$. In the following prove the conclusion one by one according to all the types of four-qubit states under LOCC.
	
($C_1$) The representative state of the $L_{0_{3\oplus\bar{1}}0_{3\oplus\bar{1}}}$ family is
$L_{0_{3\oplus\bar{1}}0_{3\oplus\bar{1}}}=|0000\rangle+|0111\rangle$ \cite{nine}.
Taking $|\psi_1 \rangle= \frac{1}{\sqrt{2}}L_{0_{3\oplus\bar{1}}0_{3\oplus\bar{1}}}$, we have
\begin{equation*}
C_1 (|\psi_1 \rangle)=0,~ C_2 (|\psi_1 \rangle) =C_3 (|\psi_1 \rangle)=C_4 (|\psi_1 \rangle) =1.
\end{equation*}
That is to say, one of $\{u$, $v$, $w\}$ is 0.
Hence, $|\psi_1 \rangle$ cannot gives rise to a tetrahedron. It is a product state of the one-qubit state $|0\rangle$ and the three-qubit $|GHZ\rangle $ state.
	
($C_2$) The representative state of the $L_{0_{7\oplus\bar{1}}}$ family is
$L_{0_{7\oplus\bar{1}}}=|0000\rangle +|1011\rangle +|1101\rangle+|1110\rangle$.
Taking $|\psi_2 \rangle= \frac{1}{2}L_{0_{7\oplus\bar{1}}}$, we have
\begin{eqnarray}
&&C_1 (|\psi_2 \rangle) =\frac{\sqrt{3}}{2},~ C_2 (|\psi_2 \rangle) =C_3 (|\psi_2 \rangle)=C_4 (|\psi_2 \rangle) =1, \nonumber\\
&&C_{12} (|\psi_2 \rangle)=C_{13} (|\psi_2 \rangle)=C_{14} (|\psi_2 \rangle)=\frac{\sqrt{5}}{2}.\nonumber
\end{eqnarray}
Therefore, $u=\frac{\sqrt{3}}{2}$, $v=w=1$. Then $G(|\psi_2 \rangle)=H(|\psi \rangle)-3R=0.92953>0$. $|\psi_2 \rangle$ can form a tetrahedrons and $|\psi_2 \rangle$ is genuine multipartite entangled state.
	
We need to consider the derivatives of $\mathcal{V}_{1234}(|\psi_2\rangle)$ with respect to $C$,
\begin{eqnarray}
&\frac{\partial \mathcal{V}_{1234}(|\psi_2\rangle)}{\partial u}=0.1049>0,~
\frac{\partial \mathcal{V}_{1234}(|\psi_2\rangle)}{\partial v}=0.0692>0,~
\frac{\partial \mathcal{V}_{1234}(|\psi_2\rangle)}{\partial w}=0.0692>0, \nonumber\\
&\frac{\partial \mathcal{V}_{1234}(|\psi_2\rangle)}{\partial C_{12}(|\psi_2\rangle)}=0.0389>0,~
\frac{\partial \mathcal{V}_{1234}(|\psi_2\rangle)}{\partial C_{13}(|\psi_2\rangle)}=0.0542>0,~
\frac{\partial \mathcal{V}_{1234}(|\psi_2\rangle)}{\partial C_{14}(|\psi_2\rangle)}=0.0387>0.\nonumber
\end{eqnarray}
Thus the monotonicity of $\mathcal{V}_{1234}(|\psi_2\rangle)$ holds and hence $\mathcal{V}_{1234}(|\psi_2\rangle)$ is non-increasing under LOCC.
	
($C_3$) The representative state of the $L_{0_{5\oplus\bar{3}}}$ family is
$L_{0_{5\oplus\bar{3}}}=|0000\rangle +|0101\rangle +|1000\rangle+|1110\rangle$.
Let $|\psi_3 \rangle=\frac{1}{2}L_{0_{5\oplus\bar{3}}}$. We have
\begin{align*}
&C_1 (|\psi_3 \rangle) =C_2 (|\psi_3 \rangle)=C_4 (|\psi_3 \rangle) =\frac{\sqrt{3}}{2},~ C_3 (|\psi_3 \rangle) =1, \\\nonumber
&C_{12} (|\psi_3 \rangle)=\frac{\sqrt{5}}{2},~ C_{13} (|\psi_3 \rangle)=C_{14} (|\psi_3 \rangle)=1.
\end{align*}
We have $u=v=w=\frac{\sqrt{3}}{2}$ and $G(|\psi_3 \rangle)=0.92298>0$.
Therefore, $|\psi_3 \rangle$ can form a tetrahedrons and $|\psi_3 \rangle$ is genuine multipartite entangled state.
	
The derivatives of $\mathcal{V}_{1234}(|\psi_3\rangle)$ with respect to $C$ are given by
\begin{eqnarray}
&\frac{\partial \mathcal{V}_{1234}(|\psi_3\rangle)}{\partial u}=0.0587>0,~
\frac{\partial \mathcal{V}_{1234}(|\psi_3\rangle)}{\partial v}=0.0783>0,~
\frac{\partial \mathcal{V}_{1234}(|\psi_3\rangle)}{\partial w}=0.0783>0, \nonumber\\
&\frac{\partial \mathcal{V}_{1234}(|\psi_3\rangle)}{\partial C_{12}(|\psi_3\rangle)}=0.0101>0,~
\frac{\partial \mathcal{V}_{1234}(|\psi_3\rangle)}{\partial C_{13}(|\psi_3\rangle)}=0.0452>0,~
\frac{\partial \mathcal{V}_{1234}(|\psi_3\rangle)}{\partial C_{14}(|\psi_3\rangle)}=0.0452>0. \nonumber
\end{eqnarray}
Thus the monotonicity of $\mathcal{V}_{1234}(|\psi_3\rangle)$ holds and hence $\mathcal{V}_{1234}(|\psi_3\rangle)$ is non-increasing under LOCC.
	
($C_4$) The representative state of the $L_{a_{2}0_{3\oplus 1}}$ family is
$L_{a_20_{3\oplus\bar{1}}}=a(|0000\rangle+|1111\rangle)+(|0011\rangle+|0101\rangle+|0110\rangle)$.
Set $|\psi_4 \rangle=\frac{1}{\sqrt{2a^2+3}}L_{a_{2}0_{3\oplus 1}}$.
We have
\begin{align*}
&C_1 (|\psi_4 \rangle) =\frac{\sqrt{4 a^2(a^2+3)}}{2 a^2+3},~ C_2 (|\psi_4\rangle)=C_3(|\psi_4\rangle)=C_4(|\psi_4\rangle)=\frac{2 \sqrt{a^4+3 a^2+2}}{2 a^2+3},\\\nonumber
&C_{12} (|\psi_4 \rangle)=C_{13} (|\psi_4 \rangle)=C_{14} (|\psi_4 \rangle)=\frac{2 \sqrt{a^4+4 a^2+2}}{2 a^2+3}.
\end{align*}
	
\begin{figure}
\includegraphics[width=18cm]{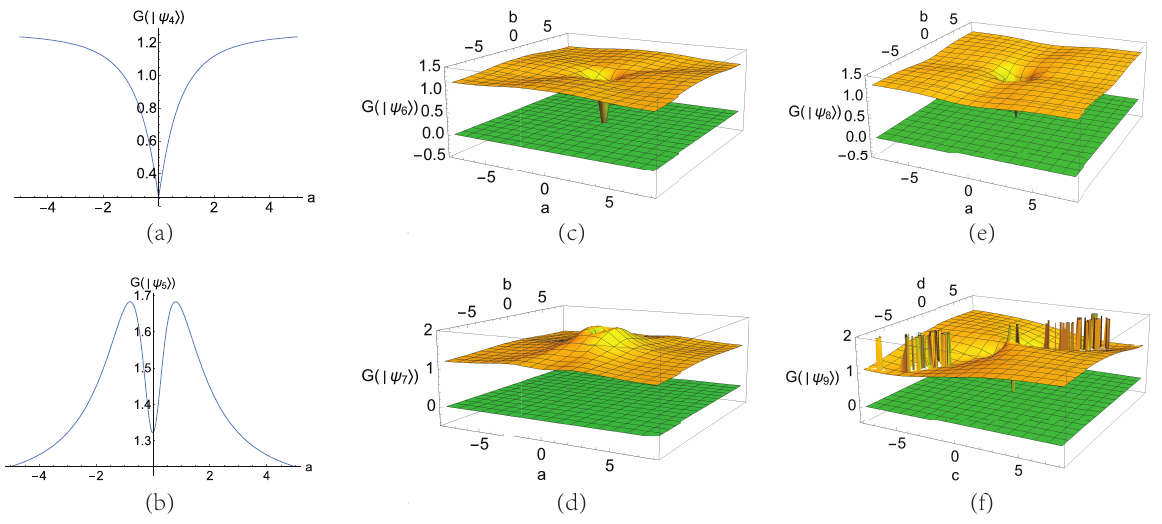}
\caption{The yellow (green) surface represents $G(|\psi \rangle)$ (zero plane) in (c)-(f). (a) $G(|\psi_4 \rangle)$ with $a \neq 0$; (b) $G(|\psi_5 \rangle)$; (c) $G(|\psi_6 \rangle)$, where $a$ and $b$ are not both 0; (d) $G(|\psi_7 \rangle)$; (e) $G(|\psi_8 \rangle)$; (f) $G(|\psi_9 \rangle)$, where $d\neq c \neq 0$.}
\label{Fig2}
\end{figure}

If $a=0$, one has $C_1 (|\psi_4 \rangle)=0$ and $C_2 (|\psi_4\rangle)=C_3(|\psi_4\rangle)=C_4(|\psi_4\rangle)=\frac{2 \sqrt{2}}{3}$. Then $u=0$ and $|\psi_4 \rangle$ cannot form a tetrahedron. It is a product state of the one-qubit state $|0\rangle$ and the three-qubit $|W\rangle$ state.
When $a \neq 0$, $C_1(|\psi_4 \rangle)>0$, $C_2 (|\psi_4\rangle)=C_3(|\psi_4\rangle)=C_4(|\psi_4\rangle)>0$ and $\frac{\sqrt{4 a^2(a^2+3)}}{2 a^2+3}<\frac{2 \sqrt{a^4+3 a^2+2}}{2 a^2+3}$. So, $u=\frac{\sqrt{4 a^2(a^2+3)}}{2 a^2+3}$ and $v=w=\frac{2 \sqrt{a^4+3 a^2+2}}{2 a^2+3}$. Then we have
\begin{align*}
H(|\psi_4 \rangle)&=u+v+w=2 \sqrt{\frac{a^2 \left(a^2+3\right)}{\left(2 a^2+3\right)^2}}+4 \sqrt{\frac{a^4+3 a^2+2}{\left(2 a^2+3\right)^2}},\nonumber\\
G(|\psi_4 \rangle)&=-\frac{2 \sqrt{3} \left(\frac{a^4+4 a^2+2}{\left(2 a^2+3\right)^2}\right)^{\frac{3}{2}}}{\sqrt{\frac{\left(a^4+4 a^2+2\right)^2}{\left(2 a^2+3\right)^4}}}+2 \sqrt{\frac{a^2 \left(a^2+3\right)}{\left(2 a^2+3\right)^2}}+4 \sqrt{\frac{a^4+3 a^2+2}{\left(2 a^2+3\right)^2}}.
\end{align*}
We have $G(|\psi_4 \rangle)>0$, see Fig. \ref{Fig2} (a).
Hence, if $a \neq 0$ $|\psi_4 \rangle$ can form a tetrahedrons and thus $|\psi_4 \rangle$ is genuine multipartite entangled state.
	
Next, we can get the derivatives of $\mathcal{V}_{1234}(|\psi_4\rangle)$ with respect to $C$, the expression of $\frac{\partial \mathcal{V}_{1234}(|\psi_4 \rangle)}{\partial C(|\psi\rangle)}$ is given in Appendix A. So the monotonicity of $\mathcal{V}_{1234}(|\psi_4\rangle)$ holds and hence $\mathcal{V}_{1234}(|\psi_4\rangle)$ is non-increasing under LOCC.
	
($C_5$) The representative state of the $L_{a_4}$ family is
$L{a_4}=a(|0000\rangle +|0101\rangle +|1010\rangle+|1111\rangle)+(i|0001\rangle +|0110\rangle -i|1011\rangle)$.
Let $|\psi_5 \rangle=\frac{1}{\sqrt{4a^2+3}}L_{a_4}$. We have
\begin{align*}
&C_1 (|\psi_5 \rangle) =C_2 (|\psi_5 \rangle) =\frac{\sqrt{8 \left(2 a^4+7 a^2+2\right)}}{4 a^2+3},~ C_3 (|\psi_5 \rangle) =C_4 (|\psi_5 \rangle)=\frac{\sqrt{8 \left(2 a^4+7 a^2+1\right)}}{4 a^2+3}, \nonumber\\
&C_{12} (|\psi_5 \rangle)=C_{14} (|\psi_5 \rangle)=\frac{\sqrt{4 \left(6 a^4+14 a^2+3\right)}}{4 a^2+3},~ C_{13} (|\psi_5 \rangle)=\frac{\sqrt{8 \left(6 a^2+1\right)}}{4 a^2+3}.
\end{align*}
Since $C_1 (|\psi_5 \rangle)>C_3 (|\psi_5 \rangle)$, $u=v=\frac{\sqrt{8 \left(2 a^4+7 a^2+1\right)}}{4 a^2+3}$ and $w=\frac{\sqrt{8 \left(2 a^4+7 a^2+2\right)}}{4 a^2+3}$, we have
\begin{align*}
H(|\psi_5 \rangle)&=u+v+w=2 \sqrt{2} \left(2 \sqrt{\frac{2 a^4+7 a^2+1}{\left(4 a^2+3\right)^2}}+\sqrt{\frac{2 a^4+7 a^2+2}{\left(4 a^2+3\right)^2}}\right),\nonumber\\
G(|\psi_5 \rangle)&=\sqrt{2} \left(-\frac{3 \left(6 a^4+14 a^2+3\right) \sqrt{\frac{6 a^2+1}{\left(4 a^2+3\right)^2}}}{\left(4 a^2+3\right)^2 \sqrt{\frac{72 \left(a^2+2\right) a^4+52 a^2+5}{\left(4 a^2+3\right)^4}}}+4 \sqrt{\frac{2 a^4+7 a^2+1}{\left(4 a^2+3\right)^2}}+2 \sqrt{\frac{2 a^4+7 a^2+2}{\left(4 a^2+3\right)^2}}\right).
\end{align*}
As shown in Fig. \ref{Fig2} (b), $G(|\psi_5 \rangle)>0$.
Hence $|\psi_5 \rangle$ can form a tetrahedrons and $|\psi_5\rangle$ is genuine multipartite entangled state.
	
Next, we can get the derivatives of $\mathcal{V}_{1234}(|\psi_5\rangle)$ with respect to $C$, the expression of $\frac{\partial \mathcal{V}_{1234}(|\psi_5 \rangle)}{\partial C(|\psi\rangle)}$ is given in Appendix B.
So the monotonicity of $\mathcal{V}_{1234}(|\psi_5\rangle)$ holds and hence $\mathcal{V}_{1234}(|\psi_5\rangle)$ is non-increasing under LOCC.
	
($C_6$) The representative state of the $L_{a_2b_2}$ family is
$L_{a_2b_2}=a(|0000\rangle +|1111\rangle)+b(|0101\rangle+|1010\rangle)+|0110\rangle +|0011\rangle$. Setting $|\psi_6 \rangle=\frac{1}{\sqrt{2(1+a^2+b^2)}}L_{a_2b_2}$, we have
\begin{align*}
&C_1(|\psi_6\rangle)=C_2(|\psi_6\rangle)=\frac{\sqrt{2 a^2 \left(b^2+1\right)+a^4+b^2 \left(b^2+2\right)}}{a^2+b^2+1},~ C_3 (|\psi_6 \rangle) =C_4 (|\psi_6 \rangle)=1, \nonumber\\
&C_{12}(|\psi_6\rangle)=C_{14}(|\psi_6\rangle)=\frac{\sqrt{a^2\left(4b^2+2\right)+a^4+\left(b^2+1\right)^2}}{a^2+b^2+1},~ C_{13} (|\psi_6 \rangle)=\frac{\sqrt{-2 a^2 \left(b^2-2\right)+a^4+b^2 \left(b^2+4\right)}}{a^2+b^2+1}.
\end{align*}
When $a=b=0$, we have $C_1(|\psi_6\rangle)=C_2(|\psi_6\rangle)=0$ and $u=v=0$. Hence, $|\psi_6 \rangle$ cannot form a tetrahedron, which is a product state:
$\frac{1}{\sqrt{2}}(|01\rangle _{13}\otimes (|01\rangle +|10\rangle )_{24})$.
	
Consider the case that $a$ and $b$ are not both 0. Since $0<C_1(|\psi_6\rangle)<C_3 (|\psi_6 \rangle)$, we have $u=v=\frac{\sqrt{2 a^2 \left(b^2+1\right)+a^4+b^2 \left(b^2+2\right)}}{a^2+b^2+1}$ and $w=1$. We get
\begin{align*}
H(|\psi_6 \rangle)&=u+v+w=2 \sqrt{1-\frac{1}{\left(a^2+b^2+1\right)^2}}+1 ,\nonumber\\
G(|\psi_6 \rangle)&=1-\frac{3 \left(a^2 \left(4 b^2+2\right)+a^4+\left(b^2+1\right)^2\right) \sqrt{\frac{-2 a^2 \left(b^2-2\right)+a^4+b^4+4 b^2}{\left(a^2+b^2+1\right)^2}}}{\left(a^2+b^2+1\right)^2 \sqrt{\frac{\left(-2 a^2 \left(b^2-2\right)+a^4+b^4+4 b^2\right) \left(2 a^2 \left(9 b^2+2\right)+3 a^4+3 b^4+4 b^2+4\right)}{\left(a^2+b^2+1\right)^4}}}+2 \sqrt{1-\frac{1}{\left(a^2+b^2+1\right)^2}}.
\end{align*}
The $G(|\psi_6 \rangle)$ is shown in Fig. \ref{Fig2} (c). Obviously, $G(|\psi_6 \rangle)>0$. Thus $|\psi_6 \rangle$ can form a tetrahedron and $|\psi_6 \rangle$ is genuine multipartite entangled state.
	
\begin{figure}
\includegraphics[width=16cm]{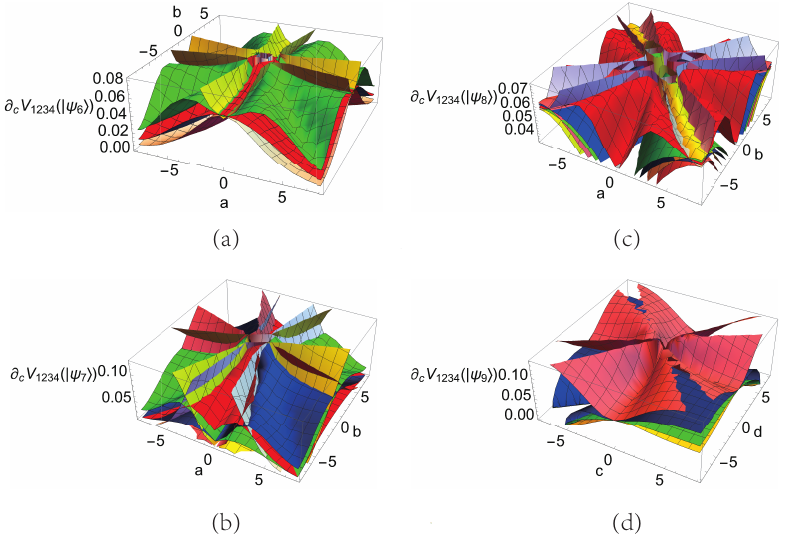}
\caption{The red surface represents $\frac{\partial \mathcal{V}_{1234}(|\psi \rangle)}{\partial u}$, the green surface represents $\frac{\partial \mathcal{V}_{1234}(|\psi \rangle)}{\partial v}$, the blue surface represents $\frac{\partial \mathcal{V}_{1234}(|\psi \rangle)}{\partial w}$, the white surface represents $\frac{\partial \mathcal{V}_{1234}(|\psi \rangle)}{\partial C_{12}(|\psi\rangle)}$, the yellow surface represents $\frac{\partial \mathcal{V}_{1234}(|\psi \rangle)}{\partial C_{13}(|\psi\rangle)}$, and the pink surface represents $\frac{\partial \mathcal{V}_{1234}(|\psi \rangle)}{\partial C_{14}(|\psi\rangle)}$.}
\label{Fig4}
\end{figure}

Next, we can get the derivatives of $\mathcal{V}_{1234}(|\psi_6\rangle)$ with respect to $C$ are shown in Fig. \ref{Fig4} (a), and the expression of $\frac{\partial \mathcal{V}_{1234}(|\psi_6 \rangle)}{\partial C(|\psi\rangle)}$ is given in Appendix C. Obviously, when $a$ and $b$ are not all 0 at the same time, $\frac{\partial \mathcal{V}_{1234}(|\psi_6\rangle)}{\partial C(|\psi_6\rangle)}\geq0$.
Thus the monotonicity of $\mathcal{V}_{1234}(|\psi_6\rangle)$ holds and hence $\mathcal{V}_{1234}(|\psi_6\rangle)$ is non-increasing under LOCC.
	
($C_7$) The representative state of $L_{ab_3}$ is
\begin{align*}
L_{ab_3}=&a(|0000\rangle+|1111\rangle)+\frac{a+b}{2}(|0101\rangle+|1010\rangle)
+\frac{a-b}{2}(|0110\rangle +|1001\rangle) \\\nonumber
&+\frac{i}{\sqrt{2}}(|0001\rangle +|0010\rangle+|0111\rangle+|1011\rangle).
\end{align*}
With normalization we set
\begin{align*}
|\psi_7 \rangle=&\frac{1}{\sqrt{3 a^2+b^2+2}}L_{ab_3}.
\end{align*}
We have
\begin{align*}
C_1 (|\psi_7 \rangle) =&C_2 (|\psi_7 \rangle)=C_3 (|\psi_7 \rangle) =C_4 (|\psi_7 \rangle) =\frac{\sqrt{a^2 \left(6 b^2+44\right)+9 a^4+b^4+12 b^2+3}}{3 a^2+b^2+2}, \\\nonumber
C_{12}(|\psi_7\rangle)=&\frac{\sqrt{4 \left(a^2 \left(3 b^2+10\right)+3 a^4+2 b^2+1\right)}}{3 a^2+b^2+2},~
C_{13}(|\psi_7\rangle)=\frac{\sqrt{M_{11}-24 a^3 b+16 a b}}{\sqrt{2} \left(3 a^2+b^2+2\right)}, \\\nonumber
C_{14}(|\psi_7\rangle)=&\frac{\sqrt{M_{11}+24 a^3 b-16 a b}}{\sqrt{2} \left(3 a^2+b^2+2\right)},
\end{align*}
and
\begin{align*}
H(|\psi_7 \rangle)&=u+v+w=3 \sqrt{\frac{a^2 \left(6 b^2+44\right)+9 a^4+b^4+12 b^2+3}{\left(3 a^2+b^2+2\right)^2}} ,\nonumber\\
G(|\psi_7 \rangle)&=\frac{12X_{11} \sqrt{\frac{8 a^2-17}{\left(3 a^2+b^2+2\right)^2}+\frac{8}{3 a^2+b^2+2}+1}-3 \sqrt{\frac{a^2 \left(6 b^2+20\right)+6 a^4+4 b^2+2}{\left(3 a^2+b^2+2\right)^2}} \sqrt{\frac{N_{11}-8 a \left(2-3 a^2\right) b}{\left(3 a^2+b^2+2\right)^2}} \sqrt{\frac{8 a \left(2-3 a^2\right) b+N_{11}}{\left(3 a^2+b^2+2\right)^2}}}{4X_{11}},
\end{align*}
where
$M_{11}=a^2 \left(6 b^2+88\right)+15 a^4+3 b^4+24 b^2+8$, $N_{11}=6 \left(a^2+4\right) b^2+15 a^4+88 a^2+3 b^4+8$, $X_{11}=\sqrt{\frac{a^6 \left(294-45 b^2\right)+a^4 \left(9 \left(b^2+42\right) b^2+707\right)+a^2 \left(9 b^6+90 b^4+322 b^2+128\right)+27 a^8+6 b^6+43 b^4+32 b^2+6}{\left(3 a^2+b^2+2\right)^4}}$.
According to Fig. \ref{Fig2} (d), we have $G(|\psi_7 \rangle)>0$. So, $|\psi_7 \rangle$ can form a triangular pyramid and is genuine multipartite entangled state.
	
The derivatives of $\mathcal{V}_{1234}(|\psi_7\rangle)$ with respect to $C$ are shown in Fig. \ref{Fig4} (b), and the expression of $\frac{\partial \mathcal{V}_{1234}(|\psi_7 \rangle)}{\partial C(|\psi\rangle)}$ is given in Appendix D. As shown in Fig. \ref{Fig4} (b), all the derivatives of $\mathcal{V}_{1234}(|\psi_7\rangle)$ with respect to $C$ satisfy $\frac{\partial \mathcal{V}_{1234}(|\psi_7\rangle)}{\partial C(|\psi_7\rangle)}>0$.
Thus the monotonicity of $\mathcal{V}_{1234}(|\psi_7\rangle)$ holds and hence $\mathcal{V}_{1234}(|\psi_7\rangle)$ is non-increasing under LOCC.
	
($C_8$) The representative state of $L_{abc_2}$ is
\begin{align*}
L_{abc_2}=&\frac{a+b}{2}(|0000\rangle+|1111\rangle)+\frac{a-b}{2}(|0011\rangle+|1100\rangle) +c(|0101\rangle +|1010\rangle)+|0110\rangle.
\end{align*}
From the normalized state of $L_{abc_2}$, $|\psi_8 \rangle= \frac{1}{\sqrt{1+a^2+b^2+2c^2}}L_{abc_2}$, we have
\begin{eqnarray}
C_1 (|\psi_8 \rangle) &=&C_2 (|\psi_8 \rangle)=C_3 (|\psi_8 \rangle) =C_4 (|\psi_8 \rangle)\nonumber\\
&=&\frac{\sqrt{2 a^2 M_{12}+2 b^2\left( c^2+N_{12}\right) +4 c^2 N_{12} +L_{12}}}{a^2+b^2+2c^2+1},\nonumber\\
C_{12}(|\psi_8\rangle)&=&\frac{\sqrt{4\left(a^2 M_{12}+b^2 \left( c^2+N_{12}\right) +c^4\right)}}{a^2+b^2+2c^2+1}, \nonumber\\
C_{13} (|\psi_8 \rangle)&=&\frac{\sqrt{O_{12}+8 a b \left(c^2+ N_{12} \right)+3 L_{12}}}{\sqrt{2} \left( a^2+b^2+2 c^2+1\right)}, \nonumber\\
C_{14} (|\psi_8 \rangle)&=&\frac{\sqrt{O_{12}-8 a b \left(c^2+ N_{12} \right)+3 L_{12}}}{\sqrt{2} \left(a^2+b^2+2 c^2+1\right)}, \nonumber
\end{eqnarray}
where $M_{12}=b^2+2c^2+1$, $N_{12}=c^2+1$, $L_{12}=a^4+b^4$, $O_{12}=2 a^2 \left( M_{12}+1\right)+4 b^2 N_{12}+8 c^2\left( N_{12}+1\right)$.
When $a=b=c=0$, $u=v=w=0$, $|\psi_8 \rangle$ cannot form a tetrahedron. It is a full separable state.
	
In \cite{Di} the authors divide the family $L_{abc_{2}}$ into three subfamilies,
the subfamily $L_{abc_{2}}$ with $c=0$, with $abc\neq 0$ and with $c\neq 0$ and $ab=0$.
Substituting these three subfamilies into $C_{i}(|\psi_8 \rangle)$ and $C_{ij}(|\psi_8 \rangle)$ respectively, we all have $C_{i}(|\psi_8 \rangle)>0$ and $G(|\psi_8 \rangle) >0$.
When $c=0$, see Fig. \ref{Fig2} (e), $|\psi_8 \rangle$ can form a triangular pyramid and is genuine multipartite entangled state.
	
Consider the derivatives of $\mathcal{V}_{1234}(|\psi_8\rangle)$ with respect to $C$ for $c=0$, and the expression of $\frac{\partial \mathcal{V}_{1234}(|\psi_8 \rangle)}{\partial C(|\psi\rangle)}$ is given in Appendix E. As shown in Fig. \ref{Fig4} (c), we have
$\frac{\partial \mathcal{V}_{1234}(|\psi_8\rangle)}{\partial C(|\psi_8\rangle)}>0$.
Similar to the case of $c=0$, one easily obtains $\frac{\partial \mathcal{V}_{1234}(|\psi_8\rangle)}{\partial C(|\psi_8\rangle)}>0$ for $abc\neq 0$ or $c\neq 0$, $ab=0$. Thus the monotonicity of $\mathcal{V}_{1234}(|\psi_8\rangle)$ holds and hence $\mathcal{V}_{1234}(|\psi_8\rangle)$ is non-increasing under LOCC.
	
($C_9$) The representative state of $G_{abcd}$ is
\begin{align*}
G_{abcd}=&\frac{a+d}{2}(|0000\rangle+|1111\rangle)+\frac{a-d}{2}(|0011\rangle+|1100\rangle)+\frac{b+c}{2}(|0101\rangle+|1010\rangle)+\frac{b-c}{2}(|0110\rangle +|1001\rangle).
\end{align*}
Set
\begin{eqnarray}
|\psi_9 \rangle=\frac{1}{\sqrt{a^2+ab+\frac{3}{2}b^2+c^2-ad+\frac{d^2}{2}}}G_{abcd}.
\end{eqnarray}
We have
\begin{eqnarray}
C_1 (| \psi_9 \rangle) &=& C_2 (|\psi_9 \rangle)=C_3 (|\psi_9 \rangle) =C_4 (|\psi_9 \rangle)\nonumber\\
&=& \frac{\sqrt{2 \left(M_{22}+N_{22}+4 a b L_{22} \right) }}{2 a^2+2 a b-2ad+3 b^2+2 c^2+d^2},\nonumber\\
C_{12}(|\psi_9 \rangle)&=&\frac{\sqrt{2 \left(M_{32}+N_{32}+4 a^2 \left(L_{22}+b^2\right)\right)}}{2 a^2+2 a b-2ad+3 b^2+2 c^2+d^2},\nonumber\\
C_{13} (|\psi_9 \rangle)&=&\frac{\sqrt{2 \left(M_{42}-N_{42}+L_{42}-24 a b c d\right) }}{2 a^2+2 a b-2ad+3 b^2+2 c^2+d^2},\nonumber\\
C_{14} (|\psi_9 \rangle)&=&\frac{\sqrt{2 \left(M_{42}-N_{42}+L_{42}+24 a b c d\right) }}{2 a^2+2 a b-2ad+3 b^2+2 c^2+d^2},\nonumber
\end{eqnarray}
where
\begin{align*}
M_{22}=&8 a^3 b+2 a^4+4 \text{ad}^2+8 b^2 c^2+7 b^4+2 c^4+2 b^2 d^2-d^4, \\\nonumber
N_{22}=&4 a^2 \left(-2 a d+3 b^2+c^2\right)-4 a d \left(3 b^2+2 c^2+d^2\right), \\\nonumber
L_{22}=&-2a d+3 b^2+2 c^2+d^2, \\\nonumber
M_{32}=&8 a^3 b+4 \text{ad}^2+12 b^2 c^2+6 b^2 d^2+5 b^4+4 c^2 d^2-3 d^4, \\\nonumber
N_{32}=&4 a b \left(-2 \text{ad}+3 b^2+2 c^2+d^2\right)-4 \text{ad} \left(3 b^2+2 c^2+d^2\right), \\\nonumber
M_{42}=&8 a^3 b+3 a^4+4 \text{ad}^2+6 b^2 c^2+8 b^4-2 c^2 d^2+3 c^4, \\\nonumber
N_{42}=&2 a^2 \left(4 \text{ad}-5 b^2-c^2+d^2\right)+4 \text{ad} \left(3 b^2+2 c^2+d^2\right), \\\nonumber
L_{42}=&4 a b \left(-2 \text{ad}+3 b^2+2 c^2+d^2\right).
\end{align*}
For the cases $a=b=c=d$; $x=y=z=0$ and $u\neq 0$; $x=y=z=-u$; $x=y=-z=-u$, where $x$, $y$, $z$, $u$ are distinct and $x,y,z,u\in \{a,b,c,d\}$,
$|\psi_9 \rangle$ cannot form a tetrahedron if $u=v=w=0$, which is a product state of two $EPR$ pairs.
	
In addition to the above special cases, the authors in \cite{Di} divide the family $G_{abcd}$ into four subfamilies: 1) $x=y=0$ and $zu\neq 0$, where $x \neq y \neq u\neq v\in \{a$, $b$, $c$, $d\}$; 2) $x=\pm y\neq 0$ and $u=\pm v\neq 0$, where $x \neq y \neq u\neq v\in \{a$, $b$, $c$, $d\}$; 3) $a=\pm d\neq 0$ and $b\neq \pm c $, or $b=\pm c\neq 0$ and $a\neq \pm d$; 4)
$x\neq \pm y$, or $x\neq \pm y$\ but only one $r=s$, where $x$, $y\in \{a$, $b$, $c$, $d\}$, $r\in \{\pm a$, $\pm d\}$ and $s\in \{\pm b$, $\pm c\}$. By straightforward calculation we have $C_{i}(|\psi_9 \rangle)>0$ and $G(|\psi_9 \rangle) >0$ for all these subcases. Fig. \ref{Fig2} (f) shows that case of $x=\pm y\neq 0$, $u=\pm v\neq 0$ in $a=-d$ and $b=c$. Hence, $|\psi_9 \rangle$ can form a triangular pyramid and is genuine multipartite entangled state.
	
Let $x=\pm y\neq 0$, $u=\pm v\neq 0$ in $a=-d$ and $b=c$, we consider the derivatives of $\mathcal{V}_{1234}(|\psi_9\rangle)$ with respect to $C$, the expression of $\frac{\partial \mathcal{V}_{1234}(|\psi_9 \rangle)}{\partial C(|\psi\rangle)}$ is given in Appendix F.
We easily get that, see Fig. \ref{Fig4} (d),
$\frac{\partial \mathcal{V}_{1234}(|\psi_9\rangle)}{\partial C(|\psi_9\rangle)}\geq0$.
Similarly, we can get that for the remaining subfamilies of $G_{abcd}$, $\frac{\partial \mathcal{V}_{1234}(|\psi_9\rangle)}{\partial C(|\psi_8\rangle)}\geq 0$ also holds.
Thus the monotonicity of $\mathcal{V}_{1234}(|\psi_9\rangle)$ holds and hence $\mathcal{V}_{1234}(|\psi_9\rangle)$ is non-increasing under LOCC.
	
The above $a,b,c,d$ are all the unique eigenvalues of a $2n\times 2n$ complex symmetric matrix $P$ with non-negative real part. The indices $L_{\alpha\beta\cdots}$ are representative for the Jordan block structure of $P$ (e.g. $L_{a_20_{3\oplus\bar{1}}}$ means that the eigenstructure of $P$ consists of two $2\times 2$ Jordan blocks with eigenvalues $a$ and $-a$, and a degenerated pair of dimension 3 and 1 respectively). From the above analysis on the concurrence tetrahedron with respect to the nine different types of entanglements for four-qubit systems under LOCC, the concurrence tetrahedron is a well-defined measure of genuine multipartite entanglement. $\Box$
	
Next, we compare the tetrahedron concurrence with GMC introduced in \cite{Ma}.
For $n$-partite pure states $|\Psi\rangle \in \mathcal{H}_{1}\otimes \mathcal{H}_{2}\otimes\cdots\otimes \mathcal{H}_{n}$ with $dim(\mathcal{H}_{i})=d_{i}$, $i=1,2, \cdots ,n$, the GMC is given by $C_{GME}(\ket{\Psi}):=\min\limits_{\gamma_i \in \gamma} \sqrt{2(1-\T (\rho^{2}_{A_{\gamma_i}}))}\ $,
where $\gamma=\{\gamma_i\}$ represents the set of all possible bipartitions $\{A_i|B_i\}$ of $\{1,2,\ldots,n\}$.
Consider the states $|\psi_{A} \rangle \in C_2$ and $|\psi_{B} \rangle \in C_3$,
\begin{eqnarray}
|\psi_A \rangle&=& \frac{1}{2}(|0000\rangle +|1011\rangle +|1101\rangle+|1110\rangle),\nonumber\\
|\psi_B \rangle&=&\frac{1}{2}(|0000\rangle +|0101\rangle +|1000\rangle+|1110\rangle).\nonumber
\end{eqnarray}
Direct calculation shows that $C_\text{GME}(|\psi_A\rangle)=C_\text{GME}(|\psi_B\rangle)=0.8660$, while
$\mathcal{V}_{1234}(|\psi_A\rangle)=0.1254$, $ \mathcal{V}_{1234}(|\psi_B\rangle)=0.0960$.
To further illustrate the advantage of the tetrahedron concurrence, let us choose the states $|\psi_{C} \rangle \in C_4$ and $|\psi_{D} \rangle \in C_5$,
\begin{eqnarray}
|\psi_C \rangle&=&\frac{1}{\sqrt{5}}(|0000\rangle+|1111\rangle
+|0011\rangle+|0101\rangle+|0110\rangle),\nonumber\\
|\psi_D \rangle&=&\frac{1}{\sqrt{4 \left(\frac{5 \sqrt{113}}{32}+\frac{51}{32}\right)+3}}\Big(\sqrt{\frac{5 \sqrt{113}}{32}+\frac{51}{32}}(|0000\rangle +|0101\rangle +|1010\rangle+|1111\rangle)\nonumber\\
&&+(i|0001\rangle +|0110\rangle -i|1011\rangle)\Big).\nonumber\\
\end{eqnarray}
We have $C_\text{GME}(|\psi_C\rangle)=C_\text{GME}(|\psi_D\rangle)=0.8000$, $\mathcal{V}_{1234}(|\psi_C\rangle)=0.1084$ and $ \mathcal{V}_{1234}(|\psi_D\rangle)=0.1624$.
Clearly, in both cases, the GMC cannot tell the difference between the entanglements of $|\psi_A\rangle$ and $|\psi_B\rangle$, as well as $|\psi_C\rangle$ and $|\psi_D\rangle$, since $C_\text{GME}$ only depends on the length of the shortest edge which is the same for both states. However, our concurrence tetrahedron takes into account all the information about the length of the edges. The entanglement of $|\psi_A\rangle$ detected by the concurrence tetrahedron is larger than the entanglement of $|\psi_B\rangle$, and that of $|\psi_C\rangle$ is less than that of $|\psi_D\rangle$. In this sense, GMC and concurrence tetrahedron are two inequivalent measures. The concurrence tetrahedron shows more advantages in characterizing the entanglement of four-qubit systems.
	
\section{Conclusion}
Quantum entanglement plays an crucial role in quantum information theory. Proper GME measures to quantify the genuine four-qubit entanglement faithfully are of great significance. We have constructed a concurrence tetrahedron based on the concurrence of nine different types of four-qubit states under LOCC, and presented a new measure of genuine multipartite entanglement for four-qubit systems. Furthermore, if a pure state is not GME, from our measure one can certify which part is separated from the rest. A specific example shows that compared with the GMC, our measure can characterize the genuine four-qubit entanglement in a finer way. Our approach may highlight investigations on quantification of genuine entanglement for multipartite systems with high dimensions.
	
\section{Acknowledgements}
This work is supported by the National Natural Science Foundation of China (NSFC) under Grants 12075159, 12171044 and 12175147; Beijing Natural Science Foundation (Grant No. Z190005); the Academician Innovation Platform of Hainan Province.
	
\section*{Appendix}
\subsection{The expression of $\frac{\partial \mathcal{V}_{1234}(|\psi_4 \rangle)}{\partial C(|\psi\rangle)}$}
\begin{eqnarray}
&\frac{\partial \mathcal{V}_{1234}(|\psi_4\rangle)}{\partial u}=\frac{\sqrt{\frac{a^2 \left(a^2+3\right)}{\left(2 a^2+3\right)^2}} \left(a^4+4 a^2+2\right) \left(a^4+4 a^2+6\right)}{3 \left(2 a^2+3\right)^4 \sqrt{\frac{a^2 \left(a^4+4 a^2+2\right) \left(2 a^6+13 a^4+26 a^2+18\right)}{\left(2 a^2+3\right)^6}}}\geq0,\nonumber\\
&\frac{\partial \mathcal{V}_{1234}(|\psi_4\rangle)}{\partial v}=\frac{\partial \mathcal{V}_{1234}(|\psi_4\rangle)}{\partial w}=\frac{a^2 \left(a^2+4\right) \left(a^4+4 a^2+2\right) \sqrt{\frac{1}{4}-\frac{1}{4 \left(2 a^2+3\right)^2}}}{3 \left(2 a^2+3\right)^4 \sqrt{\frac{a^2 \left(a^4+4 a^2+2\right) \left(2 a^6+13 a^4+26 a^2+18\right)}{\left(2 a^2+3\right)^6}}}\geq0,\nonumber\\
&\frac{\partial \mathcal{V}_{1234}(|\psi_4\rangle)}{\partial C_{12}(|\psi_4\rangle)}=\frac{a^2 \sqrt{\frac{a^4+4 a^2+2}{\left(2 a^2+3\right)^2}} \left(a^6+6 a^4+12 a^2+12\right)}{3 \left(2 a^2+3\right)^4 \sqrt{\frac{a^2 \left(a^4+4 a^2+2\right) \left(2 a^6+13 a^4+26 a^2+18\right)}{\left(2 a^2+3\right)^6}}}\geq0,\nonumber\\
&\frac{\partial \mathcal{V}_{1234}(|\psi_4\rangle)}{\partial C_{13}(|\psi_4\rangle)}=\frac{\partial \mathcal{V}_{1234}(|\psi_4\rangle)}{\partial C_{14}(|\psi_4\rangle)}=\frac{a^2 \left(a^2+2\right) \left(\frac{a^4+4 a^2+2}{\left(2 a^2+3\right)^2}\right)^{\frac{3}{2}}}{3 \left(2 a^2+3\right)^2 \sqrt{\frac{a^2 \left(a^4+4 a^2+2\right) \left(2 a^6+13 a^4+26 a^2+18\right)}{\left(2 a^2+3\right)^6}}}\geq0,\nonumber
\end{eqnarray}
	
\subsection{The expression of $\frac{\partial \mathcal{V}_{1234}(|\psi_5 \rangle)}{\partial C(|\psi\rangle)}$}
\begin{eqnarray}
&\frac{\partial \mathcal{V}_{1234}(|\psi_5\rangle)}{\partial v}=\frac{8 a^2 \sqrt{\frac{2 a^4+7 a^2+2}{\left(4 a^2+3\right)^2}} \left(18 a^4+27 a^2+4\right)}{3 \left(4 a^2+3\right)^4 \sqrt{\frac{\left(6 a^2+1\right) \left(60 a^8+344 a^6+496 a^4+176 a^2+15\right)}{\left(4 a^2+3\right)^6}}}>0,\nonumber\\
&\frac{\partial \mathcal{V}_{1234}(|\psi_5\rangle)}{\partial u}=\frac{\partial \mathcal{V}_{1234}(|\psi_5\rangle)}{\partial w}=\frac{2 \sqrt{\frac{2 a^4+7 a^2+1}{\left(4 a^2+3\right)^2}} \left(36 a^6+90 a^4+44 a^2+5\right)}{3 \left(4 a^2+3\right)^4 \sqrt{\frac{\left(6 a^2+1\right) \left(60 a^8+344 a^6+496 a^4+176 a^2+15\right)}{\left(4 a^2+3\right)^6}}}>0,\nonumber\\
&\frac{\partial \mathcal{V}_{1234}(|\psi_5\rangle)}{\partial C_{12}(|\psi_5\rangle)}=\frac{\partial \mathcal{V}_{1234}(|\psi_5\rangle)}{\partial C_{14}(|\psi_5\rangle)}=\frac{\sqrt{2} \left(6 a^2+1\right) \sqrt{\frac{6 a^4+14 a^2+3}{\left(4 a^2+3\right)^2}} \left(2 \left(a^2+7\right) a^2+3\right)}{3 \left(4 a^2+3\right)^4 \sqrt{\frac{\left(6 a^2+1\right) \left(60 a^8+344 a^6+496 a^4+176 a^2+15\right)}{\left(4 a^2+3\right)^6}}}>0,\nonumber\\
&\frac{\partial \mathcal{V}_{1234}(|\psi_5\rangle)}{\partial C_{13}(|\psi_5\rangle)}=\frac{\sqrt{\frac{6 a^2+1}{\left(4 a^2+3\right)^2}} \left(4 \left(15 a^6+74 a^4+80 a^2+25\right) a^2+7\right)}{3 \left(4 a^2+3\right)^4 \sqrt{\frac{\left(6 a^2+1\right) \left(60 a^8+344 a^6+496 a^4+176 a^2+15\right)}{\left(4 a^2+3\right)^6}}}>0, \nonumber
\end{eqnarray}
	
\subsection{The expression of $\frac{\partial \mathcal{V}_{1234}(|\psi_6 \rangle)}{\partial C(|\psi\rangle)}$}
\begin{eqnarray}
&&\frac{\partial \mathcal{V}_{1234}(|\psi_6\rangle)}{\partial u}=\frac{n_{11}\sqrt{1-\frac{1}{\left(a^2+b^2+1\right)^2}}}{12 \left(a^2+b^2+1\right)^4 \sqrt{\frac{m_{11}}{\left(a^2+b^2+1\right)^6}}},\nonumber\\
&&\frac{\partial \mathcal{V}_{1234}(|\psi_6\rangle)}{\partial v}=\frac{\left(10 a^2 b^2+a^4+b^4+3\right) \left(-2 a^2 \left(b^2-2\right)+a^4+b^4+4 b^2\right) \sqrt{1-\frac{1}{\left(a^2+b^2+1\right)^2}}}{12 \left(a^2+b^2+1\right)^4 \sqrt{\frac{m_{11}}{\left(a^2+b^2+1\right)^6}}} ,\nonumber\\
&&\frac{\partial \mathcal{V}_{1234}(|\psi_6\rangle)}{\partial w}=\frac{\left(-2 a^2 \left(b^2-2\right)+a^4+b^4+4 b^2-2\right) \left(a^2 \left(4 b^2+2\right)+a^4+\left(b^2+1\right)^2\right)}{12 \left(a^2+b^2+1\right)^4 \sqrt{\frac{m_{11}}{\left(a^2+b^2+1\right)^6}}},\nonumber\\
&&\frac{\partial \mathcal{V}_{1234}(|\psi_6\rangle)}{\partial C_{12}(|\psi_6\rangle)}=\frac{r_{11}\sqrt{\frac{a^2 \left(4 b^2+2\right)+a^4+\left(b^2+1\right)^2}{\left(a^2+b^2+1\right)^2}}}{12 \left(a^2+b^2+1\right)^4 \sqrt{\frac{m_{11}}{\left(a^2+b^2+1\right)^6}}},\nonumber\\
&&\frac{\partial \mathcal{V}_{1234}(|\psi_6\rangle)}{\partial C_{13}(|\psi_6\rangle)}=\frac{s_{11}\sqrt{\frac{-2 a^2 \left(b^2-2\right)+a^4+b^4+4 b^2}{\left(a^2+b^2+1\right)^2}}}{12 \left(a^2+b^2+1\right)^4 \sqrt{\frac{m_{11}}{\left(a^2+b^2+1\right)^6}}}, \nonumber\\
&&\frac{\partial \mathcal{V}_{1234}(|\psi_6\rangle)}{\partial C_{14}(|\psi_6\rangle)}=\frac{t_{11}\sqrt{\frac{a^2 \left(4 b^2+2\right)+a^4+\left(b^2+1\right)^2}{\left(a^2+b^2+1\right)^2}}}{12 \left(a^2+b^2+1\right)^4 \sqrt{\frac{m_{11}}{\left(a^2+b^2+1\right)^6}}}.\nonumber
\end{eqnarray}
where
\begin{eqnarray}
m_{11}&=&2 a^{10} \left(6 b^2+7\right)+a^8 \left(-6 b^4+94 b^2+31\right)+2 a^6 \left(-8 b^6+66 b^4+79 b^2+17\right),\nonumber\\
&&+a^4 \left(-6 b^8+132 b^6+262 b^4+70 b^2+23\right)+2 a^2 \left(6 b^{10}+47 b^8+79 b^6+35 b^4+22 b^2-1\right),\nonumber\\
&&+2 a^{12}+b^2 \left(b^2+2\right) \left(b^2 \left(b^2+4\right) \left(2 \left(b^4+b^2\right)+3\right)-1\right)-1,\nonumber\\
n_{11}&=&2 a^6 \left(b^2+3\right)+2 a^4 \left(-3 b^4+9 b^2+5\right)+2 a^2 \left(b^6+9 b^4+12 b^2+2\right)+a^8+b^8+6 b^6+10 b^4+4 b^2+2,\nonumber\\
r_{11}&=&-2 a^6 \left(b^2-3\right)+2 a^4 \left(b^4+b^2+4\right)+2 a^2 \left(-b^6+b^4+11 b^2-1\right)+a^8+b^8+6 b^6+8 b^4-2 b^2+1,\nonumber\\
s_{11}&=&16 a^6 b^2+a^4 \left(26 b^4+24 b^2-1\right)+2 a^2 \left(8 b^6+12 b^4-2 b^2+3\right)+a^8+b^8-b^4+6 b^2,\nonumber\\
t_{11}&=&-2 a^6 \left(b^2-3\right)+a^4 \left(2 \left(b^4+b^2\right)+7\right)-2 a^2 \left(b^6-b^4-6 b^2+1\right)+a^8+b^8+6 b^6+7 b^4-2 b^2-2.\nonumber
\end{eqnarray}
	
\subsection{The expression of $\frac{\partial \mathcal{V}_{1234}(|\psi_7 \rangle)}{\partial C(|\psi\rangle)}$}
\begin{eqnarray}
&&\frac{\partial \mathcal{V}_{1234}(|\psi_7\rangle)}{\partial u}=\frac{n_{22} \sqrt{\frac{8 a^2-17}{\left(3 a^2+b^2+2\right)^2}+\frac{8}{3 a^2+b^2+2}+1}}{3 \left(3 a^2+b^2+2\right)^4 \sqrt{\frac{m_{22}}{\left(3 a^2+b^2+2\right)^6}}},\nonumber\\
&&\frac{\partial \mathcal{V}_{1234}(|\psi_7\rangle)}{\partial v}=\frac{r_{22} \sqrt{\frac{8 a^2-17}{\left(3 a^2+b^2+2\right)^2}+\frac{8}{3 a^2+b^2+2}+1}}{6 \left(3 a^2+b^2+2\right)^4 \sqrt{\frac{m_{22}}{\left(3 a^2+b^2+2\right)^6}}},\nonumber\\
&&\frac{\partial \mathcal{V}_{1234}(|\psi_7\rangle)}{\partial w}=\frac{s_{22} \sqrt{\frac{8 a^2-17}{\left(3 a^2+b^2+2\right)^2}+\frac{8}{3 a^2+b^2+2}+1}}{6 \left(3 a^2+b^2+2\right)^4 \sqrt{\frac{m_{22}}{\left(3 a^2+b^2+2\right)^6}}},\nonumber\\
&&\frac{\partial \mathcal{V}_{1234}(|\psi_7\rangle)}{\partial C_{12}(|\psi_7\rangle)} =\frac{t_{22} \sqrt{\frac{a^2 \left(3 b^2+10\right)+3 a^4+2 b^2+1}{\left(3 a^2+b^2+2\right)^2}}}{24 \left(3 a^2+b^2+2\right)^4 \sqrt{\frac{m_{22}}{\left(3 a^2+b^2+2\right)^6}}},\nonumber\\
&&\frac{\partial \mathcal{V}_{1234}(|\psi_7\rangle)}{\partial C_{13}(|\psi_7\rangle)}=\frac{x_{22} \sqrt{\frac{6 \left(a^2+4\right) b^2+8 \left(2-3 a^2\right) a b+15 a^4+88 a^2+3 b^4+8}{\left(3 a^2+b^2+2\right)^2}}}{6 \sqrt{2} \left(3 a^2+b^2+2\right)^4 \sqrt{\frac{m_{22}}{\left(3 a^2+b^2+2\right)^6}}}, \nonumber\\
&&\frac{\partial \mathcal{V}_{1234}(|\psi_7\rangle)}{\partial C_{14}(|\psi_7\rangle)}=\frac{y_{22}\sqrt{\frac{6 \left(a^2+4\right) b^2+8 \left(3 a^2-2\right) a b+15 a^4+88 a^2+3 b^4+8}{\left(3 a^2+b^2+2\right)^2}}}{6 \sqrt{2} \left(3 a^2+b^2+2\right)^4 \sqrt{\frac{m_{22}}{\left(3 a^2+b^2+2\right)^6}}}.\nonumber
\end{eqnarray}
where
\begin{eqnarray}
m_{22}&=&9 a^{10} \left(2278-159 b^2\right)+27 a^8 \left(-18 b^4+594 b^2+3869\right)+2 a^6 \left(117 b^6+7638 b^4+75426 b^2+89216\right)\nonumber\\
&&+a^4 \left(369 b^8+8316 b^6+64954 b^4+134208 b^2+40200\right)+a^2 \Big(45 b^{10}+1278 b^8+10628 b^6+34432 b^4\nonumber\\
&&+20656 b^2+3136\Big)+1269 a^{12}+\left(2 b^2+1\right) \left(15 b^8+304 b^6+1416 b^4+672 b^2+80\right),\nonumber\\
n_{22}&=&\left(a^2 \left(3 b^2+10\right)+3 a^4+2 b^2+1\right) \left(3 \left(-2 a^2 \left(b^2-8\right)+a^4+b^4\right)+16 b^2+4\right) ,\nonumber\\
r_{22}&=&\left(6 \left(a^2+4\right) b^2+8 \left(2-3 a^2\right) a b+15 a^4+88 a^2+3 b^4+8\right) \left(a^2 \left(3 b^2+10\right)+6 a^3 b+3 a^4-4 a b+2 b^2+1\right),\nonumber\\
s_{22}&=&\left(6 \left(a^2+4\right) b^2+8 \left(3 a^2-2\right) a b+15 a^4+88 a^2+3 b^4+8\right) \left(a^2 \left(3 b^2+10\right)-6 a^3 b+3 a^4+4 a b+2 b^2+1\right),\nonumber\\
t_{22}&=&36 a^6 \left(3 b^2+52\right)-2 a^4 \left(87 b^4+456 b^2-4636\right)+4 a^2 \left(15 b^6+204 b^4+1556 b^2+288\right)\nonumber\\
&&-9 a^8+15 b^8+272 b^6+1016 b^4+384 b^2+32,\nonumber\\
x_{22}&=&3 a^6 \left(39 b^2+158\right)+72 a^5 b \left(b^2+10\right)+a^4 \left(57 b^4+558 b^2+913\right)+8 a^3 b \left(3 b^2 \left(b^2+8\right)-62\right)\nonumber\\
&&+a^2 \left(3 b^6+118 b^4+446 b^2+128\right)+144 a^7 b+63 a^8-16 a b \left(b^4+10 b^2+2\right)+\left(2 b^2+1\right) \left(b^4+24 b^2+4\right),\nonumber\\
y_{22}&=&3 a^6 \left(39 b^2+158\right)-72 a^5 b \left(b^2+10\right)+a^4 \left(57 b^4+558 b^2+913\right)-8 a^3 b \left(3 b^2 \left(b^2+8\right)-62\right)\nonumber\\
&&+a^2 \left(3 b^6+118 b^4+446 b^2+128\right)-144 a^7 b+63 a^8+16 a b \left(b^4+10 b^2+2\right)+\left(2 b^2+1\right) \left(b^4+24 b^2+4\right).\nonumber
\end{eqnarray}
	
\subsection{The expression of $\frac{\partial \mathcal{V}_{1234}(|\psi_8 \rangle)}{\partial C(|\psi\rangle)}$}
\begin{eqnarray}
&&\frac{\partial \mathcal{V}_{1234}(|\psi_8\rangle)}{\partial u}=\frac{\left(-2 a^2 b^2+3 a^4+3 b^4\right) \left(a^2 \left(b^2+1\right)+b^2\right) \sqrt{1-\frac{1}{\left(a^2+b^2+1\right)^2}}}{3 \left(a^2+b^2+1\right)^4 \sqrt{\frac{m_{33}}{\left(a^2+b^2+1\right)^6}}} ,\nonumber\\
&&\frac{\partial \mathcal{V}_{1234}(|\psi_8\rangle)}{\partial v}=\frac{\left(a^2 \left(b^2+1\right)-2 a b+b^2\right) \sqrt{1-\frac{1}{\left(a^2+b^2+1\right)^2}} \left(2 a^2 \left(b^2+2\right)+3 a^4+8 a b+b^2 \left(3 b^2+4\right)\right)}{6 \left(a^2+b^2+1\right)^4 \sqrt{\frac{m_{33}}{\left(a^2+b^2+1\right)^6}}} ,\nonumber\\
&&\frac{\partial \mathcal{V}_{1234}(|\psi_8\rangle)}{\partial w}=\frac{\left(a^2 \left(b^2+1\right)+2 a b+b^2\right) \sqrt{1-\frac{1}{\left(a^2+b^2+1\right)^2}} \left(2 a^2 \left(b^2+2\right)+3 a^4-8 a b+b^2 \left(3 b^2+4\right)\right)}{6 \left(a^2+b^2+1\right)^4 \sqrt{\frac{m_{33}}{\left(a^2+b^2+1\right)^6}}} \nonumber\\
&&\frac{\partial \mathcal{V}_{1234}(|\psi_8\rangle)}{\partial C_{12}(|\psi_8\rangle)} = \frac{n_{33} \sqrt{\frac{a^2 \left(b^2+1\right)+b^2}{\left(a^2+b^2+1\right)^2}}}{24 \left(a^2+b^2+1\right)^4 \sqrt{\frac{m_{33}}{\left(a^2+b^2+1\right)^6}}},\nonumber\\
&&\frac{\partial \mathcal{V}_{1234}(|\psi_8\rangle)}{\partial C_{13}(|\psi_8\rangle)}=\frac{\Big(r_{33}+a^6 \left(b^2+1\right)+a^4 \left(6 b^4+11 b^2+4\right)\Big) \sqrt{\frac{2 a^2 \left(b^2+2\right)+3 a^4+8 a b+b^2 \left(3 b^2+4\right)}{\left(a^2+b^2+1\right)^2}}}{6 \sqrt{2} \left(a^2+b^2+1\right)^4 \sqrt{\frac{m_{33}}{\left(a^2+b^2+1\right)^6}}} , \nonumber\\
&&\frac{\partial \mathcal{V}_{1234}(|\psi_8\rangle)}{\partial C_{14}(|\psi_8\rangle)}=\frac{\Big(s_{33}+a^6 \left(b^2+1\right)+a^4 \left(6 b^4+11 b^2+4\right)\Big) \sqrt{\frac{2 a^2 \left(b^2+2\right)+3 a^4-8 a b+b^2 \left(3 b^2+4\right)}{\left(a^2+b^2+1\right)^2}}}{6 \sqrt{2} \left(a^2+b^2+1\right)^4 \sqrt{\frac{m_{33}}{\left(a^2+b^2+1\right)^6}}} .\nonumber
\end{eqnarray}
where
\begin{eqnarray}
m_{33}&=&15 a^{10} \left(b^2+1\right)+a^8 \left(36 b^4+91 b^2+40\right)+2 a^6 \left(13 b^6+59 b^4+24 b^2+8\right)+2 a^4 b^2 \Big(18 b^6\nonumber\\
&&+59 b^4+40 b^2-8\Big)+a^2 b^4 \left(15 b^6+91 b^4+48 b^2-16\right)+b^6 \left(15 b^4+40 b^2+16\right) ,\nonumber\\
n_{33}&=&4 a^6 \left(5 b^2+6\right)-2 a^4 \left(3 \left(b^2+4\right) b^2+8\right)+4 a^2 b^2 \left(5 b^4-6 b^2+8\right)+15 a^8+b^4 \left(15 b^4+24 b^2-16\right),\nonumber\\
r_{33}&=&-8 a^3 \left(b^3+b\right)+a^2 b^2 \left(b^4+11 b^2+8\right)-8 a^5 b-8 a \left(b^5+b^3\right)+b^4 \left(b^2+4\right) ,\nonumber\\
s_{33}&=&8 a^3 \left(b^3+b\right)+a^2 b^2 \left(b^4+11 b^2+8\right)+8 a^5 b+8 a \left(b^5+b^3\right)+b^4 \left(b^2+4\right) ,\nonumber
\end{eqnarray}
	
\subsection{The expression of $\frac{\partial \mathcal{V}_{1234}(|\psi_9 \rangle)}{\partial C(|\psi\rangle)}$}
\begin{eqnarray}
&&\frac{\partial \mathcal{V}_{1234}(|\psi_9\rangle)}{\partial u}=\frac{\partial \mathcal{V}_{1234}(|\psi_9\rangle)}{\partial v}=\frac{n_{44} \sqrt{\frac{38 c^2 d^2-20 c^3 d+17 c^4-4 (5 c+1) d^3+13 d^4}{\left(5 c^2-2 c d+5 d^2\right)^2}}}{6 \sqrt{2} \left(5 c^2-2 c d+5 d^2\right)^4 \sqrt{\frac{m_{44}}{\left(5 c^2-2 c d+5 d^2\right)^6}}} ,\nonumber\\
&&\frac{\partial \mathcal{V}_{1234}(|\psi_9\rangle)}{\partial w}=\frac{r_{44} \sqrt{\frac{38 c^2 d^2-20 c^3 d+17 c^4-4 (5 c+1) d^3+13 d^4}{\left(5 c^2-2 c d+5 d^2\right)^2}}}{6 \sqrt{2} \left(5 c^2-2 c d+5 d^2\right)^4 \sqrt{\frac{m_{44}}{\left(5 c^2-2 c d+5 d^2\right)^6}}} \nonumber\\
&&\frac{\partial \mathcal{V}_{1234}(|\psi_9\rangle)}{\partial C_{12}(|\psi_9\rangle)} =\frac{\partial \mathcal{V}_{1234}(|\psi_9\rangle)}{\partial C_{13}(|\psi_9\rangle)}=\frac{s_{44} \sqrt{\frac{54 c^2 d^2-20 c^3 d+17 c^4-4 (5 c+1) d^3+13 d^4}{\left(5 c^2-2 c d+5 d^2\right)^2}}}{6 \sqrt{2} \left(5 c^2-2 c d+5 d^2\right)^4 \sqrt{\frac{m_{44}}{\left(5 c^2-2 c d+5 d^2\right)^6}}} ,\nonumber\\
&&\frac{\partial \mathcal{V}_{1234}(|\psi_9\rangle)}{\partial C_{14}(|\psi_9\rangle)}=\frac{t_{44} \sqrt{\frac{6 c^2 d^2-20 c^3 d+17 c^4-4 (5 c+1) d^3+13 d^4}{\left(5 c^2-2 c d+5 d^2\right)^2}}}{6 \sqrt{2} \left(5 c^2-2 c d+5 d^2\right)^4 \sqrt{\frac{m_{44}}{\left(5 c^2-2 c d+5 d^2\right)^6}}}.\nonumber
\end{eqnarray}
where
\begin{eqnarray}
m_{44}&=&\left(6 c^2 d^2-20 c^3 d+17 c^4-4 (5 c+1) d^3+13 d^4\right) \Big(2236 c^6 d^2-8 (355 c+17) c^4 d^3+2 (1887 c+80) c^3 d^4\nonumber\\
&&-8 (335 c+54) c^2 d^5-680 c^7 d+289 c^8-104 (5 c+1) d^7+4 (c (451 c+40)+4) d^6+169 d^8\Big) ,\nonumber\\
n_{44}&=&\left(6 c^2 d^2-20 c^3 d+17 c^4-4 (5 c+1) d^3+13 d^4\right) \left(54 c^2 d^2-20 c^3 d+17 c^4-4 (5 c+1) d^3+13 d^4\right) ,\nonumber\\
r_{44}&=&\left(6 c^2 d^2-20 c^3 d+17 c^4-4 (5 c+1) d^3+13 d^4\right) \left(102 c^2 d^2-20 c^3 d+17 c^4-4 (5 c+1) d^3+13 d^4\right),\nonumber\\
s_{44}&=&\left(6 c^2 d^2-20 c^3 d+17 c^4-4 (5 c+1) d^3+13 d^4\right) \left(22 c^2 d^2-20 c^3 d+17 c^4-4 (5 c+1) d^3+13 d^4\right)  ,\nonumber\\
t_{44}&=&3324 c^6 d^2-8 (515 c+17) c^4 d^3+2 (3039 c+80) c^3 d^4-8 (495 c+86) c^2 d^5-680 c^7 d+289 c^8\nonumber\\
&&-104 (5 c+1) d^7+4 (c (659 c+40)+4) d^6+169 d^8.\nonumber
\end{eqnarray}

\end{document}